%% file: main.tex
\PassOptionsToPackage{table}{xcolor}
\documentclass{article}
\usepackage{report}

\usepackage[utf8]{inputenc}
\usepackage[T1]{fontenc}
\usepackage{graphicx}
\usepackage{booktabs}
\usepackage{multirow}
\usepackage{array}
\usepackage{tabularx}
\usepackage{adjustbox}
\usepackage{amsmath,amssymb}
\usepackage{enumitem}
\usepackage{caption}
\usepackage{subcaption}
\usepackage{float}
\usepackage{wrapfig}
\usepackage{needspace}
\usepackage{tcolorbox}
\usepackage{xspace}
\usepackage{pifont}
\usepackage{makecell}
\tcbuselibrary{breakable}

\definecolor{njuPurple}{RGB}{220,205,230}
\definecolor{njuPurpleLight}{RGB}{250,245,252}
\definecolor{tableHead}{RGB}{238,230,245}
\definecolor{tableBand}{RGB}{245,241,249}
\definecolor{tableGroup}{RGB}{226,216,236}

\newtcolorbox{abstractbox}{
    colback=njuPurpleLight,
    colframe=njuPurple,
    boxrule=1pt,
    arc=4mm,
    left=8pt,
    right=8pt,
    top=8pt,
    bottom=8pt,
    opacityback=0.95,
    breakable
}

\newcommand{\bench}{OmniHalluc-L\xspace}
\newcommand{\method}{MPRC\xspace}

\newcommand{\cmark}{\checkmark}
\newcommand{\xmark}{$\times$}
\newcommand{\tblsmall}{\small\setlength{\tabcolsep}{4.2pt}\renewcommand{\arraystretch}{1.12}}
\newcommand{\tblfoot}{\footnotesize\setlength{\tabcolsep}{3.5pt}\renewcommand{\arraystretch}{1.08}}
\newcolumntype{Y}{>{\raggedright\arraybackslash}X}
\title{OmniHalluc-L: Counterfactual Benchmarking and\\
Modality-Perturbation Reliability Calibration for\\
Long-Form Omni Hallucination}

\author{Zixuan Dong*$^{,1}$ \quad Jiafu Tang*$^{,2}$ \quad Zhide Lei$^{2}$ \quad Zhe Cao$^{2}$ \quad Zijie Zhang$^{2}$ \quad \\Yanghai Wang$^{2}$ \quad Shihao Li$^{2}$ \quad Xiaodong Wang$^{1}$ \quad Baoyun Peng$^{1}$ \quad Jiaheng Liu$^\dagger$$^{,2}$\\
$^{1}$National University of Defense Technology, 
$^{2}$Nanjing University\\
\texttt{dongzixuan18@nudt.edu.cn} \quad \texttt{liujiaheng@nju.edu.cn} \\
\texttt{https://github.com/ZexDong/omnihallucL-mprc}}
\begin{document}

\maketitle

\input{content/0_Abstract}
\let\oldthefootnote\thefootnote

\let\thefootnote\relax\footnotetext{*~Equal Contribution. ~~$^\dagger$~Corresponding Author.}
\let\thefootnote\oldthefootnote
\input{content/1_MainText}

\bibliographystyle{unsrtnat}
\bibliography{references}

\appendix
\input{content/2_Appendix}

\end{document}

%% file: content/0_Abstract.tex
\begin{abstractbox}
\begin{center}
\textbf{\Large Abstract}
\end{center}
Long-video Omni assistants often fail not by inventing content, but by misbinding real evidence: they hear the right utterance and see the right event, yet attach it to the wrong speaker, moment, or modality.
These \emph{almost-true} errors evade standard video QA because local evidence remains valid, so item-level scoring can reward both a supported claim and its near-counterfactual.
We introduce a counterfactual event-binding protocol that constructs paired supported/counterfactual claims from the same audio-visual event evidence and evaluates them by strict-pair accuracy.
We instantiate it as \bench, a benchmark for long-video Omni hallucination, with 3{,}600 single-claim QA items from 638 long-form videos averaging 24.16 minutes and covering 256.87 hours.
Under this protocol, open-weight Omni models remain weak at pair-level binding: Qwen2.5-Omni-7B reaches 32.06\% and Qwen3-Omni-Instruct reaches 41.55\%, versus 76.54\% for a closed-source reference.
To narrow this gap without updating the backbone, we propose \method, Modality-Perturbation Reliability Calibration, a frozen-backbone framework that selects audio-negative probes within video-level folds and fuses their response shifts with native audio-visual confidence into per-claim support estimates.
\method lifts Qwen2.5-Omni-7B to 36.22\% and Qwen3 to 51.09\% on \bench, and improves target-adapted MCQ accuracy on OmniVideoBench ($+$2.20) and WorldSense ($+$1.51) with Qwen3.
\end{abstractbox}

%% file: content/1_MainText.tex
\section{Introduction}

\noindent
\begin{minipage}[t]{0.505\textwidth}
\vspace{0pt}
Omni assistants are moving from curated short clips to minutes-long audio-visual streams such as lectures, podcasts, and livestreams.
At this horizon, hallucination often means misbinding real evidence rather than inventing content: a model may hear the right utterance and see the right event, yet assign them to the wrong speaker, moment, scene, or modality.
The evidence is present, but the claimed relation is false.
Such \emph{almost-true} errors evade standard video QA because item-level scoring can reward both a supported claim and a minimally edited near-counterfactual.

Current evaluations ask adjacent questions.
Long-video benchmarks measure comprehension and reasoning \citep{videomme2025,longvideobench2024,mlvu2024}, hallucination benchmarks have moved from images to vision-only video settings \citep{pope2023,videohallucer2024,elvhalluc2025}, and Omni evaluations target broad audio-visual understanding and real-world omnimodal reasoning \citep{omnivideobench2026,worldsense2026,lvomnibench2026,mmou2026}.
What remains missing is whether recurring evidence is attached to the right source, time, scene, and modality in a long stream.
\end{minipage}
\hfill
\begin{minipage}[t]{0.465\textwidth}
\vspace{0pt}
\centering
\includegraphics[width=\linewidth]{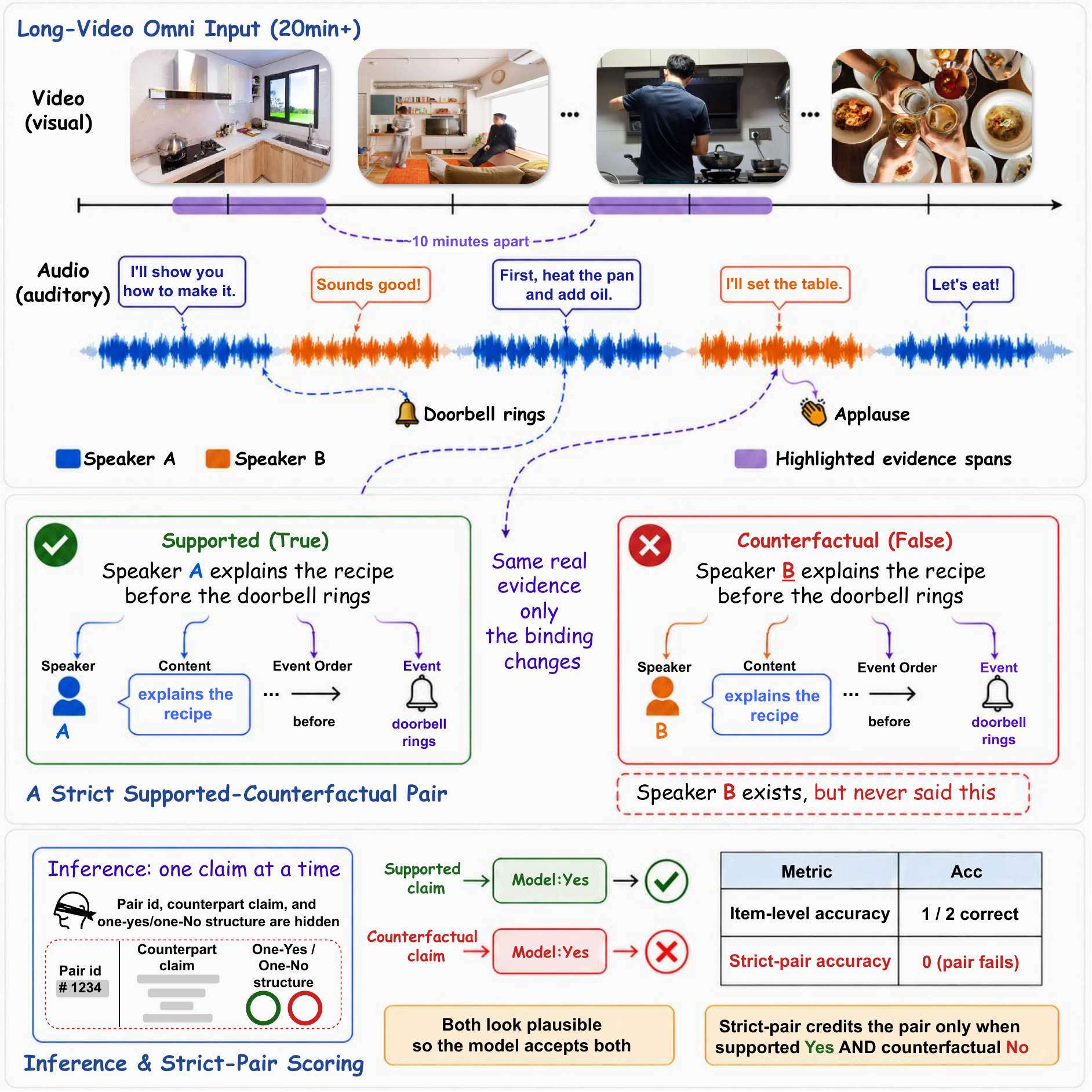}
\captionof{figure}{Strict-pair scoring: paired claims share real evidence but differ in binding.}
\label{fig:motivation}
\end{minipage}

\vspace{0.35em}

We address this gap with a counterfactual event-binding protocol.
Starting from real audio-visual event evidence, we author a supported claim and a minimally edited counterfactual that preserves the evidence but changes its binding relation.
Pair identity is hidden: models verify one claim at a time, and strict-pair accuracy credits only accepting the supported claim and rejecting the counterfactual.
This turns hallucination evaluation from local plausibility checking into relation-consistency testing.

We instantiate this protocol as \bench, a benchmark for long-video Omni hallucination.
\bench contains 3{,}600 single-claim QA items from 638 long-form videos averaging 24.16 minutes and totaling 256.87 hours, organized into 1{,}800 hidden supported/counterfactual pairs across three evidence-binding families (Section~\ref{sec:bench}).
The same view motivates \method (\textbf{M}odality-\textbf{P}erturbation \textbf{R}eliability \textbf{C}alibration), a frozen-backbone framework that treats structured audio-negative branches as reliability probes and calibrates their response shifts with native audio-visual confidence.

Experiments reveal a large reliability gap and an actionable path to reduce it.
Qwen2.5-Omni-7B reaches 32.06\% SPA and Qwen3-Omni-Instruct reaches 41.55\%, while the strongest closed-source reference reaches 76.54\%.
Without updating the Omni backbone, \method lifts them to 36.22\% and 51.09\%, outperforming the MAD contrastive decoder \citep{mad2026}.
The same audio-counterfactual signal improves target-adapted MCQ accuracy on OmniVideoBench~\citep{omnivideobench2026} ($+$2.20) and WorldSense~\citep{worldsense2026} ($+$1.51) with Qwen3-Omni-Instruct.

Our contributions are: (i) a counterfactual audio-visual event-binding construction and strict-pair evaluation protocol; (ii) \bench, a 3{,}600-item long-video Omni hallucination benchmark built under this protocol; (iii) \method, a frozen-backbone reliability calibration framework based on structured audio-negative probes; and (iv) a systematic study of open-weight, frozen-decoding, and closed-source Omni systems under strict evidence-binding evaluation.

\section{Related Work}

\paragraph{Long-video understanding and Omni evaluation.}
Long-video benchmarks test retrieval, summarization, event localization, and reasoning over extended temporal context \citep{egoschema2023,longvideobench2024,mlvu2024,videomme2025,longvale2025}.
Omni, speech-centric, and inconsistency benchmarks evaluate audio-visual understanding, real-world omnimodal reasoning, speaker-level audiovisual grounding, and injected conflicts \citep{omnivideobench2026,worldsense2026,avspeakerbench2025,lvomnibench2026,mmou2026,avid2026}.
They cover comprehension, perception, and inconsistency handling broadly, but not the stricter question of whether recurring real evidence is bound to the correct speaker, moment, scene, or modality under hidden supported/counterfactual pairs.

\paragraph{Multimodal hallucination benchmarks.}
Hallucination benchmarks began with image-level object and language-vision conflicts \citep{pope2023,hallusionbench2024,amber2023,mmhalbench2023} and later expanded to video event, temporal, semantic aggregation, and compositional failures \citep{videohallucer2024,eventhallusion2024,vidhalluc2025,elvhalluc2025,omnicvhall2026}.
AVHBench and AV-ConfuseBench further test cross-modal hallucination, matching, reasoning, and confusion from modified or absent sounds \citep{avhbench2025,avconfusebench2026}.
These resources expose fragile audio-visual relations, but \bench targets long-form cases where both claims reuse real evidence and only the binding relation is false.

\paragraph{Frozen-model hallucination mitigation.}
Training-free mitigation methods use decoding-time calibration or contrastive signals without updating the backbone \citep{cad2023,lcd2024,vcd2024,opera2024}.
Audio-visual decoders such as MAD and AVCD compare native responses with perturbed modality views \citep{mad2026,avcd2025}.
These methods are useful frozen-decoding references; however, long-form binding errors are heterogeneous, so \method estimates reliability over structured probes rather than relying on one universal perturbation rule.

\begin{figure}[!tbp]
  \centering
  \includegraphics[width=0.92\textwidth]{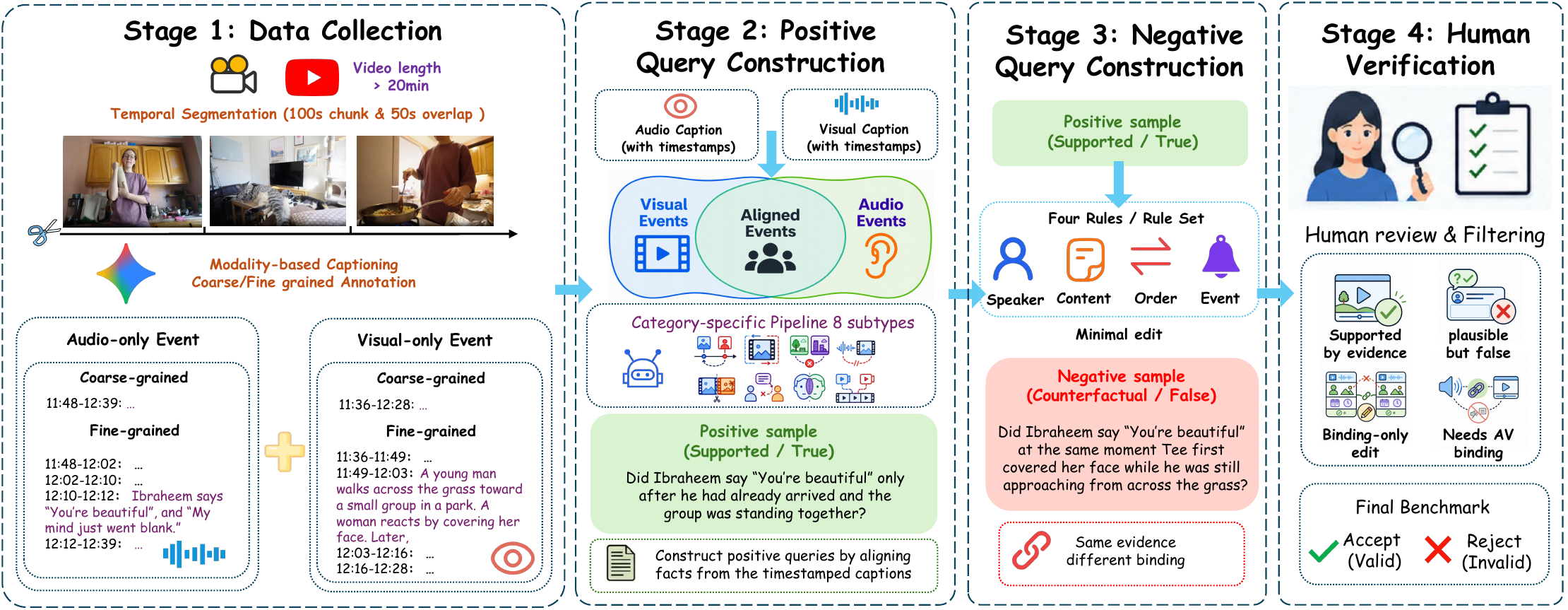}
  \caption{\bench construction pipeline. Long-form videos are temporally segmented, annotated through isolated audio and visual views, authored into supported/counterfactual claim pairs, and validated by human verification so that each pair differs only in the evidence-binding relation.}
  \label{fig:benchmark_overview}
\end{figure}

\section{The \bench Benchmark}
\label{sec:bench}

\subsection{Counterfactual Claim Verification}

\bench evaluates whether an Omni model binds real audio-visual evidence to the correct relation.
Given a long video and a natural-language claim, the model answers \textsc{Yes} or \textsc{No}.
Each instance belongs to a hidden strict pair $p=(v_p,q_p^{+},q_p^{-})$: $q_p^{+}$ is supported by the video, while $q_p^{-}$ preserves the same evidence but rebinds it to the wrong speaker, scene, moment, or modality.
At test time, the model receives only one claim; the pair identifier, paired counterfactual, and one-positive-per-pair structure are hidden.
The task therefore tests whether a model can reject a false relation over real evidence, not merely whether it can detect that relevant evidence exists.

\begin{figure}[!tbp]
\centering
\captionsetup{font=small,skip=3pt}

\begin{minipage}[t]{0.535\linewidth}
  \vspace{0pt}
  \centering
  \tblfoot
  \renewcommand{\arraystretch}{1.08}
  \setlength{\tabcolsep}{4.0pt}
  \begin{adjustbox}{width=\linewidth}
  \begin{tabular}{@{}lcccc@{}}
  \toprule
  \rowcolor{tableHead}
  \textbf{Dataset} & \textbf{Long} & \textbf{Omni} & \textbf{Hall.} & \textbf{SPA} \\
  \midrule
  VideoHallucer   & \xmark & \xmark & \cmark & \xmark \\
  EventHallusion  & \xmark & \xmark & \cmark & \xmark \\
  VidHalluc       & \xmark & \xmark & \cmark & \xmark \\
  AVHBench        & \xmark & \cmark & \cmark & \xmark \\
  AV-ConfuseBench & \xmark & \cmark & \cmark & \xmark \\
  ELV-Halluc      & \cmark & \xmark & \cmark & \xmark \\
  OmniVCHall      & \xmark & \xmark & \cmark & \xmark \\
  AVID            & \xmark & \cmark & \cmark & \xmark \\
  \midrule
  \rowcolor{tableBand}
  \textbf{\bench} & \cmark & \cmark & \cmark & \cmark \\
  \bottomrule
  \end{tabular}
  \end{adjustbox}
  \captionof{table}{Benchmark coverage comparison.}
  \label{tab:comparison}
\end{minipage}
\hfill
\begin{minipage}[t]{0.425\linewidth}
  \vspace{0pt}
  \centering
  \includegraphics[
    width=\linewidth,
    trim={1pt 1pt 1pt 1pt},
    clip
  ]{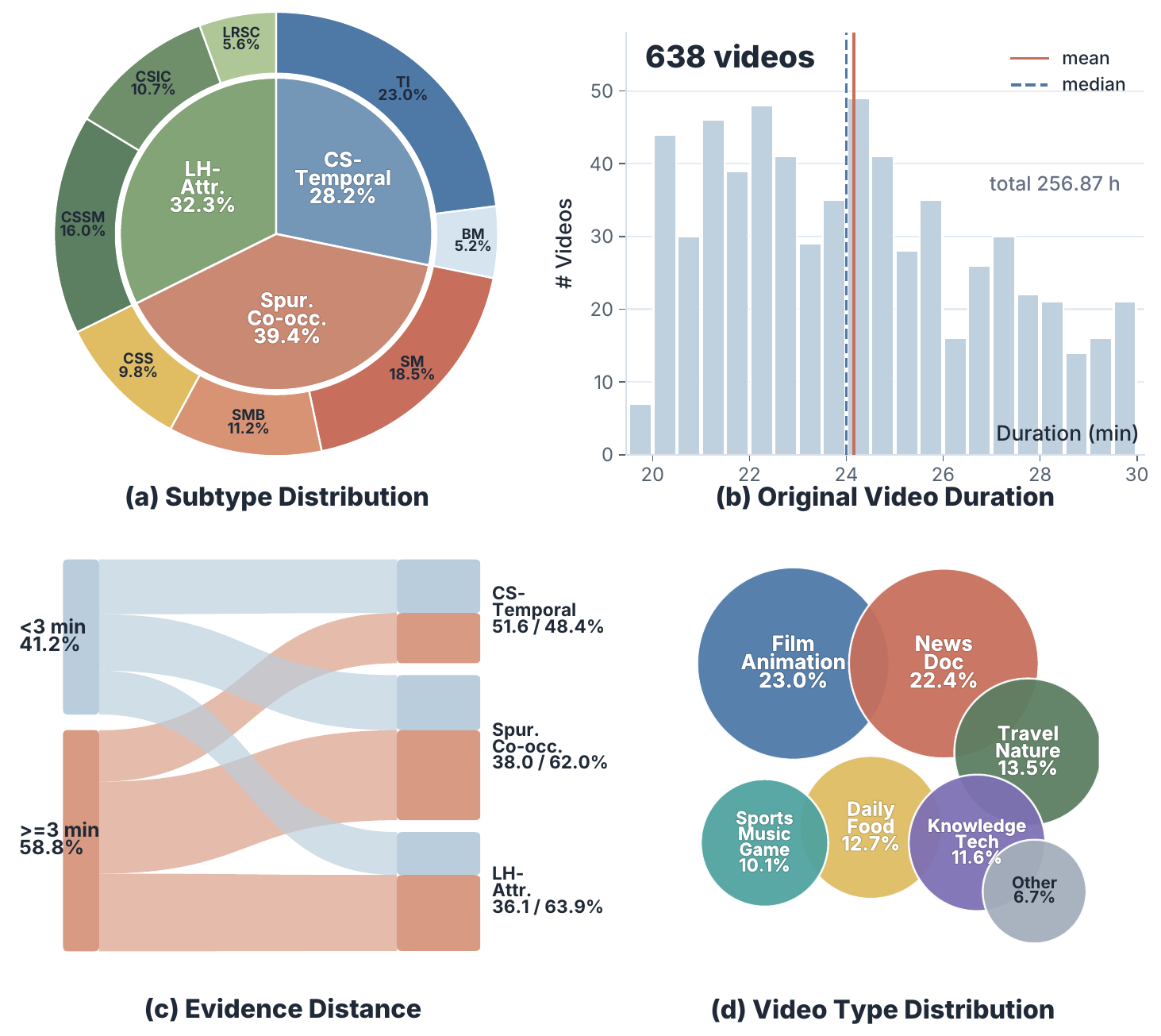}
  \captionof{figure}{\bench dataset statistics.}
  \label{fig:statistics}
\end{minipage}

\vspace{0.15em}
\begin{minipage}{0.975\linewidth}
\footnotesize
\textit{Notes.}
For Table~\ref{tab:comparison}, Long denotes average duration above 5 minutes; Omni denotes joint audio-visual input; Hall. denotes hallucination, AV-confusion, or cross-modal inconsistency evaluation; SPA denotes strict hidden supported/counterfactual pair scoring.
For Figure~\ref{fig:statistics}, panels summarize subtype and family distributions, source-video duration, evidence distance, and video type coverage over 3{,}600 verified QA items from 638 long-form videos.
\end{minipage}

\vspace{-0.45em}
\end{figure}

A pair is credited only when both sides are answered correctly. The primary metric is strict-pair accuracy:
\begin{equation}
\label{eq:spa}
\begin{aligned}
\mathrm{SPA}
&=
\frac{1}{|\mathcal{P}|}
\sum_{p\in\mathcal{P}}
\mathbb{I}\!\left[g(v_p,q_p^{+})=1\right]
\cdot
\mathbb{I}\!\left[g(v_p,q_p^{-})=0\right].
\end{aligned}
\end{equation}
SPA is 0\% for constant \textsc{Yes}/\textsc{No} predictors and 25\% in expectation for random guessing, making answer-prior shortcuts visible rather than hiding them inside item-level accuracy.
Table~\ref{tab:comparison} situates this protocol against vision-only hallucination, audio-visual hallucination or confusion, and injected-inconsistency benchmarks.

\subsection{Benchmark Construction}

\bench is built from 638 long-form source videos averaging 24.16 minutes and totaling 256.87 hours, a regime where speakers, sounds, scenes, and events recur across many minutes.
We segment each video into overlapping 100-second windows with a 50-second stride, limiting caption drift while preserving boundary-crossing speech and audio-visual cues.

Construction follows the same principle as the evaluation protocol: keep the evidence real and vary only the binding relation.
We first use modality-isolated annotation, describing audio without visual context and visual content without audio context, so the pipeline does not pre-solve cross-modal grounding.
Category-specific authoring protocols then produce a supported claim and a near-counterfactual that changes only the relation being tested.
Human verification checks entailment, false-but-plausible counterfactuality, binding-only edits, and shortcut leakage; additional verification and normalization details are summarized in Appendix~\ref{sec:appendix_benchmark_details}.

The verified benchmark contains 3{,}600 single-claim QA items across three relation-defined families.
\textbf{Cross-Stream Temporal} tests timing and boundary alignment between audio and visual streams.
\textbf{Spurious Co-occurrence} tests false association among real scene, sound, and action evidence.
\textbf{Long-Horizon Attribution} tests speaker attribution, identity continuity, and source tracking across distant moments.
Together, they cover the timing, co-occurrence, and attribution relations that long-video Omni models must preserve.
This relation-centric taxonomy makes near-counterfactual evaluation possible: the local evidence remains real, while the tested claim changes only how that evidence is connected.

Evidence intervals are retained as diagnostic metadata but are not provided to models.
The evaluated task remains single-query verification over the model's normalized video input; subtype definitions, evidence metadata, and verification details are summarized in Appendix~\ref{sec:appendix_benchmark_details}.

\subsection{Evaluation Protocol}
Open-weight Omni models are scored from canonical \textsc{Yes}/\textsc{No} logits using a binary margin and softmax-normalized support score, while closed-source systems are parsed from generated text with bounded retries; token choices, API identifiers, decoding settings, and failure-handling rules are given in Appendix~\ref{sec:appendix_answer_normalization}.

The primary metric is SPA (Eq.~\ref{eq:spa}); item accuracy, balanced item accuracy, and predicted-\textsc{Yes} rate are reported only as diagnostics because item-level metrics can mask pair collapse.
All held-out evaluations use the same single-query setting.
For models with duration constraints, we apply fixed query-agnostic temporal preprocessing and 256-frame sampling, with model-specific sidecar or union-clip formatting where required; no evidence span, retrieval window, or pair-level information is selected per query.

\subsection{Benchmark Results}

\begin{table}[!tbp]
\centering
\tblfoot
\renewcommand{\arraystretch}{1.10}
\setlength{\tabcolsep}{3.0pt}
\begin{adjustbox}{width=\linewidth,center}
\begin{tabular}{@{}l c rr rrr rrr r@{}}
\toprule
\rowcolor{tableHead}
& & \multicolumn{2}{c}{\textit{CS-Temporal}} 
& \multicolumn{3}{c}{\textit{Spur. Co-occ.}} 
& \multicolumn{3}{c}{\textit{LH-Attr.}} & \\
\cmidrule(lr){3-4}\cmidrule(lr){5-7}\cmidrule(lr){8-10}
\rowcolor{tableHead}
\textbf{Model} & \textbf{In.} 
& \textbf{TI} & \textbf{BM} 
& \textbf{SM} & \textbf{SMB} & \textbf{CSS} 
& \textbf{CSSM} & \textbf{CSIC} & \textbf{LRSC} 
& \textbf{All} \\
\midrule

\rowcolor{tableGroup}
\multicolumn{11}{c}{\textbf{Open-weight Omni models}} \\
Qwen2.5-Omni-3B     & AV & 14.98 & 18.09 & 29.43 & 34.83 & 31.82 & 25.00 & 23.32 & 38.61 & 25.50 \\
Qwen2.5-Omni-7B     & AV & 27.05 & 14.89 & 39.04 & 38.31 & 41.48 & 22.57 & 34.72 & 38.61 & 32.06 \\
OmniVinci           & AV & 28.99 & 21.28 & 32.13 & 37.81 & 42.05 & 25.35 & 31.09 & 30.69 & 31.17 \\
MiniCPM-o~4.5       & AV & 35.75 & 26.60 & 36.04 & 41.29 & 47.73 & 24.31 & 37.82 & 48.51 & 36.22 \\
Qwen3-Omni-Instruct & AV & 37.92 & 37.23 & 43.07 & 38.19 & 44.83 & 40.28 & 48.44 & 47.00 & 41.55 \\
\midrule

\rowcolor{tableGroup}
\multicolumn{11}{c}{\textbf{Open-weight video-only models}} \\
Qwen2.5-VL-3B       & V  & 14.49 & 18.09 & 28.23 & 32.34 & 35.80 & 21.18 & 26.94 & 28.71 & 24.50 \\
Qwen2.5-VL-7B       & V  & 19.57 & 19.15 & 28.53 & 31.34 & 30.68 & 19.10 & 38.34 & 33.66 & 26.33 \\
Qwen3-VL-32B        & V  & 28.02 & 20.21 & 29.82 & 33.17 & 41.95 & 18.06 & 46.88 & 37.00 & 30.79 \\
\midrule

\rowcolor{tableGroup}
\multicolumn{11}{c}{\textbf{Closed-source Omni models}} \\
Gemini 3.1 Flash-Lite & AV & 50.00 & 38.30 & 63.83 & 62.44 & 55.43 & 61.97 & 69.63 & 50.00 & 57.86 \\
Gemini 3 Flash        & AV & 65.45 & 55.32 & 74.09 & 61.31 & 69.14 & 65.61 & 81.25 & 60.00 & 67.83 \\
Gemini 3.1 Pro        & AV & 81.31 & 68.09 & 78.55 & 74.75 & 73.71 & 73.59 & 82.29 & 64.00 & 76.54 \\
\bottomrule
\end{tabular}
\end{adjustbox}

\caption{Strict-pair accuracy (\%) on the full \bench benchmark. All models follow the same single-query protocol; pair identity is hidden at test time.}
\label{tab:benchmark_leaderboard}
\end{table}
Table~\ref{tab:benchmark_leaderboard} reports SPA on the full \bench benchmark.
Current open-weight Omni systems remain far below the strongest closed-source reference: Qwen2.5-Omni-7B reaches 32.06\% SPA and Qwen3-Omni-Instruct reaches 41.55\%, while Gemini 3.1 Pro reaches 76.54\%.
The gap leaves substantial headroom, while nontrivial video-only baselines and stronger matched Omni results show that \bench is not reducible to language plausibility or visual recognition alone.
Errors span all three binding families, making \bench a targeted diagnostic for long-form evidence binding rather than a generic video QA leaderboard.

\section{\method: Modality-Perturbation Reliability Calibration}
\label{sec:method}
Long-video hallucination is difficult to mitigate because the relevant evidence may already be present in the model's response pathway.
The failure is often relational: a real utterance is assigned to the wrong speaker, a real sound is attached to the wrong scene, or a true event is carried beyond the moment that supports it.
A single audio perturbation cannot diagnose all such failures.
Temporal shifts stress cross-stream alignment, and segment swaps stress local co-occurrence and source continuity.
The same perturbation, however, can be informative for one claim and irrelevant for another.
\method therefore treats audio perturbations not as a universal contrastive recipe, but as reliability probes for estimating whether the model's native binding judgment should be trusted.
The probe is not meant to replace the original evidence; it asks whether the model's support decision remains coherent when the audio relation most relevant to binding is disturbed.

\begin{figure}[H]
  \centering
  \includegraphics[width=0.99\textwidth]{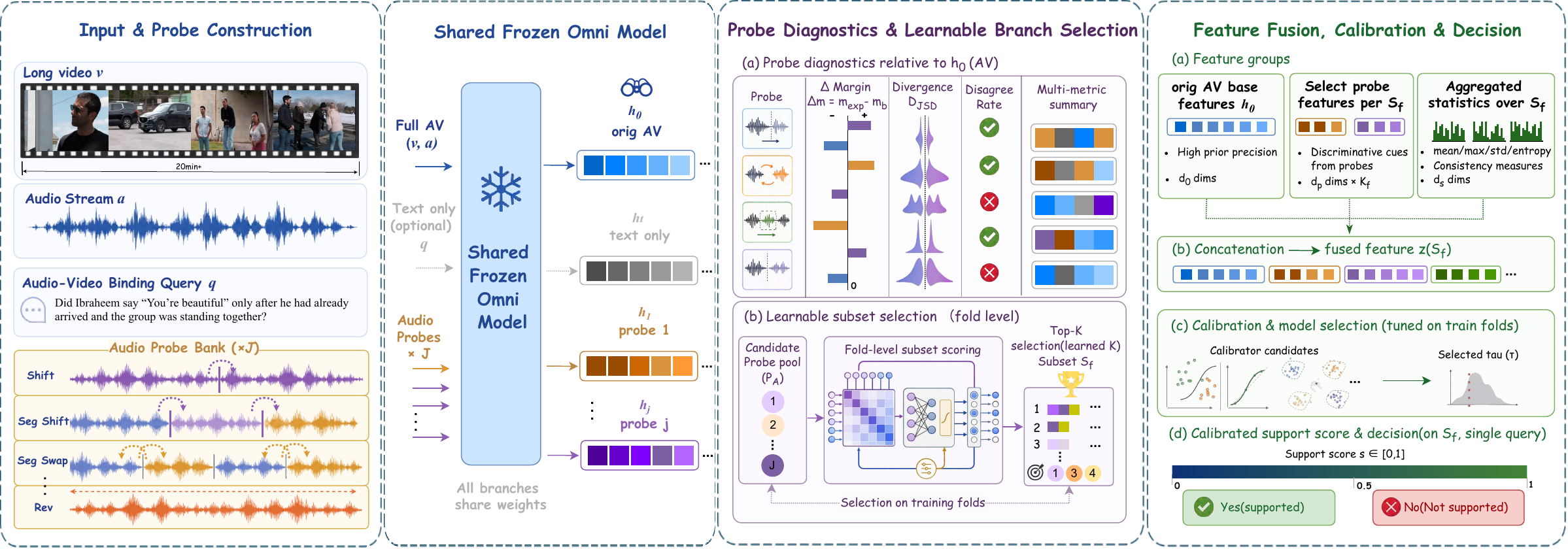}
  \caption{\method overview. A frozen Omni backbone is queried under native audio-visual input and structured audio-negative probes. Branch subsets and a lightweight calibrator are selected within video-level folds; held-out claims are predicted independently from native confidence and branch-induced response shifts.}
  \label{fig:mprc}
\end{figure}

\method is a frozen-backbone calibration framework that keeps the visual stream fixed, queries native audio-visual and structured audio-negative views, and converts response shifts into support features.
A lightweight reliability layer, selected within the training folds, combines native confidence with selected audio-negative signals without updating the backbone.
This reframes mitigation as post-hoc evidence-binding calibration: the model supplies perceptual and sensitivity signals, while \method learns when those signals justify per-claim support.

\paragraph{Reliability features.}
Let $F$ be the frozen Omni backbone and let $(v,q)$ denote a video-claim instance.
The native audio-visual query produces $h_0 = F(v,q)$.
From $h_0$, we extract $\phi_{\mathrm{av}}(h_0)$, including the parsed answer, support confidence, Yes/No margin, and response uncertainty when available.
For an audio-negative branch $b$, let $T_b(v)$ denote the same visual stream paired with a perturbed audio stream, and let $h_b = F(T_b(v),q)$.
The branch feature $\phi_b(h_b,h_0)$ contains the branch answer and confidence as well as deltas in support score, margin, and uncertainty relative to the native AV view.
We also query a text-only probe $F(q)$ to expose language-prior behavior; the released main configuration includes this block and reports a no-text ablation.

For a candidate branch subset $\mathcal{B}$, the feature representation is
\begin{equation}
\label{eq:mprc_features}
\begin{aligned}
\mathbf{x}_{\mathcal{B}}(v,q)
&=
\Big[
\phi_{\mathrm{av}}(h_0),\;
\phi_{\mathrm{text}}(F(q),h_0),
\{\phi_b(h_b,h_0)\}_{b\in\mathcal{B}}
\Big].
\end{aligned}
\end{equation}
The representation remains claim-sensitive through $(v,q)$-dependent response shifts, while the branch subset is fixed by training-video selection and never chosen from test-time pair information.

\paragraph{Audio-negative probe bank.}
The probe bank is designed to disturb audio-visual binding while preserving the visual evidence.
Short global shifts and segment-level shifts probe \textbf{Cross-Stream Temporal} errors by changing the relative timing between speech, sounds, and visual events.
Adjacent segment swaps probe \textbf{Spurious Co-occurrence} errors by keeping plausible local audio content while attaching it to a neighboring visual context.
No branch is assumed to be universally helpful.
The released compact branch pool contains \texttt{sh30}, \texttt{sh60}, \texttt{seg60}, \texttt{seg90}, and \texttt{swap60}; branch subsets are selected on training videos and listed in Appendix~\ref{sec:appendix_audio_bank}.

\paragraph{Calibration and model selection.}
For each video-level outer fold $f$, \method searches over branch subsets, thresholds, and lightweight heads including logistic-regression variants, ridge classifiers, and gradient-boosted trees, using only training videos.
Each head maps $\mathbf{x}_{\mathcal{B}}(v_i,q_i)$ to an item-level support estimate $y_i\in\{0,1\}$; pair identifiers are never input features.
Model selection then uses out-of-fold training predictions to optimize strict-pair reliability:
\begin{equation}
\label{eq:mprc_selection}
\begin{aligned}
(\mathcal{B}_f^{\star},h_f^{\star},\tau_f^{\star})
&=
\arg\max_{\mathcal{B},h,\tau}^{\mathrm{lex}}
\mathbf{m}_{\mathcal{B},h,\tau}, \\
\mathbf{m}_{\mathcal{B},h,\tau}
&=
\Big[
\mathrm{SPA}^{\mathrm{OOF}},
\mathrm{Acc}_{\mathrm{item}}^{\mathrm{OOF}},
-|\hat{p}_{Y}^{\mathrm{OOF}}-0.5|,
-|\mathcal{B}|
\Big].
\end{aligned}
\end{equation}
Here $\mathrm{SPA}^{\mathrm{OOF}}$ uses training-video pair identifiers, $\mathrm{Acc}_{\mathrm{item}}^{\mathrm{OOF}}$ is item accuracy, and $\hat{p}_{Y}^{\mathrm{OOF}}$ is the predicted-\textsc{Yes} rate.
The lexicographic rule prioritizes SPA, item accuracy, balanced predicted-\textsc{Yes} rate, and compact branch subsets; ties prefer the full feature mode.
Because selection is video-level, held-out videos cannot influence branch choice, calibration, or threshold tuning.

\paragraph{Pair-blind test-time inference.}
After selection, a held-out claim $(v,q)$ in fold $f$ is scored by
\begin{equation}
\label{eq:mprc_inference}
\begin{aligned}
s_f(v,q)
&=
h_f^{\star}\!\left(\mathbf{x}_{\mathcal{B}_f^{\star}}(v,q)\right), \\
g_f(v,q)
&=
\mathbb{I}[s_f(v,q)\ge \tau_f^{\star}].
\end{aligned}
\end{equation}
At test time, \method predicts each claim independently.
It receives no pair identifier, paired counterfactual, evidence span, or query-specific retrieval window, and imposes no one-positive-per-pair constraint.
Pair identifiers are used only for training-validation model selection and final evaluation, so deployment remains single-query and pair-blind.

\paragraph{Frozen-backbone reliability layer.}
\method is deliberately post-hoc.
No gradients flow into the audio encoder, visual encoder, or language model, and no video evidence is relabeled at test time.
The method does not add new perceptual capacity; following standard post-hoc calibration practice \citep{guo2017calibration}, it calibrates whether the backbone's own evidence-binding judgment is reliable under native confidence and structured audio-negative sensitivity.
The calibration target is relational rather than merely probabilistic: branch-induced shifts test whether support remains coherent when binding-sensitive audio structure is disturbed.
This separation is useful for long-form Omni models: as the backbone improves, its confidence margins and perturbation responses can become cleaner reliability signals, while \method converts those signals into stricter support decisions without changing the model itself.

\section{Experiments}

\subsection{Experimental Protocol}

All experiments use video-level five-fold splits: every query from the same source video is assigned to the same fold, and trainable components are fit only on training videos before evaluation on held-out videos.
We report held-out fold means, with confidence intervals computed by video-level bootstrap over out-of-fold predictions.

The central comparison is backbone-training-free reliability improvement.
Raw~AV uses the frozen model, MAD is the closest frozen-decoding reference with contrastive strength selected within training, and \method fits a lightweight reliability layer on labeled training videos without updating the Omni backbone.
Because existing mitigation methods are not designed for paired long-form Omni verification, the comparison asks whether a fixed Omni model can be made more reliable through post-hoc evidence-binding calibration, not whether backbone training improves perception.
Additional details on MAD tuning and comparison scope are given in Appendix~\ref{sec:appendix_mad_scope}.

The primary metric is strict-pair accuracy (SPA).
Item accuracy, balanced item accuracy, and predicted-\textsc{Yes} rate are tracked as diagnostics because item-level metrics can hide answer priors and pair collapse.
We evaluate open-weight Omni models including Qwen2.5-Omni-7B~\citep{qwen25omni2025}, Qwen3-Omni-Instruct~\citep{qwen3omni2025}, OmniVinci~\citep{omnivinci2025}, and MiniCPM-o~4.5~\citep{minicpmo452026}.
We also include Qwen2.5-VL and Qwen3-VL as video-only references~\citep{qwen25vl2025,qwen3vl2025}.
Closed-source Omni models are reported only as Raw~AV references because their logits and confidence features are not exposed.

\begin{table}[!tbp]
  \centering
  \tblfoot
  \setlength{\tabcolsep}{3.2pt}
  \renewcommand{\arraystretch}{1.45}
  \begin{adjustbox}{max width=\textwidth}
  \begin{tabular}{@{}llcrr@{\hspace{8pt}}!{\color{gray!45}\vrule width 0.4pt}@{\hspace{8pt}}llcrr@{}}
  \toprule
  \rowcolor{tableHead}
  \textbf{Model}
  & \textbf{Method}
  & \textbf{C/S/L}
  & \textbf{All}
  & \textbf{$\Delta$}
  & \textbf{Model}
  & \textbf{Method}
  & \textbf{C/S/L}
  & \textbf{All}
  & \textbf{$\Delta$} \\
  \midrule
  \multirow{3}{*}{Qwen2.5-O-7B}
  & Raw AV  & 24.80/39.44/29.38 & 32.06$_{[29.76,34.47]}$ & --
  & \multirow{3}{*}{Qwen3-O-Inst.}
  & Raw AV  & 37.80/42.13/44.14 & 41.55$_{[39.19,44.03]}$ & -- \\
  & MAD     & 25.98/41.27/32.30 & 34.06$_{[31.80,36.42]}$ & $+$2.00
  &
  & MAD     & 40.75/46.38/45.86 & 44.62$_{[42.17,47.07]}$ & $+$3.07 \\
  & \method & \textbf{27.36/42.68/36.08} & \textbf{36.22$_{[33.82,38.57]}$} & \textbf{$+$4.16}
  &
  & \method & \textbf{43.31/52.34/56.38} & \textbf{51.09$_{[48.66,53.48]}$} & \textbf{$+$9.54} \\
  \midrule
  \multirow{3}{*}{OmniVinci}
  & Raw AV  & 27.56/36.20/28.18 & 31.17$_{[28.94,33.41]}$ & --
  & \multirow{3}{*}{MiniCPM-o~4.5}
  & Raw AV  & 34.06/40.42/32.99 & 36.22$_{[33.84,38.52]}$ & -- \\
  & MAD     & 25.79/38.31/29.55 & 31.94$_{[29.81,34.11]}$ & $+$0.77
  &
  & MAD     & 37.20/41.69/35.22 & 38.33$_{[35.90,40.84]}$ & $+$2.11 \\
  & \method & \textbf{27.95/38.56/30.29} & \textbf{32.89$_{[30.64,35.17]}$} & \textbf{$+$1.72}
  &
  & \method & \textbf{38.19/45.49/39.35} & \textbf{41.44$_{[39.10,43.83]}$} & \textbf{$+$5.22} \\
  \bottomrule
  \end{tabular}
  \end{adjustbox}
  \caption{\bench main results on the full benchmark (\%). C/S/L denotes Cross-Stream Temporal, Spurious Co-occurrence, and Long-Horizon Attribution. All reports SPA with video-level bootstrap 95\% confidence intervals as small subscripts, and $\Delta$ is over Raw~AV.}
  \label{tab:main_result}
\end{table}

\subsection{Main Results}
\label{sec:main_results}
Table~\ref{tab:main_result} shows that fixed contrastive decoding is useful but limited.
MAD improves every open-weight backbone (+0.77 to +3.07 SPA), showing that modality perturbation exposes reliability signal even without backbone updates.
Appendix~\ref{app:fixed_branch_diag} further shows that some training-free fixed branches are already competitive with MAD, but the best branch is not stable across data distributions and output formats.
\method therefore learns a post-hoc reliability layer on training videos rather than hard-coding a universal negative rule.
This raises Qwen2.5-Omni from 32.06\% to 36.22\%, Qwen3-Omni-Instruct from 41.55\% to 51.09\%, OmniVinci from 31.17\% to 32.89\%, and MiniCPM-o~4.5 from 36.22\% to 41.44\%.

For Qwen2.5-Omni, the largest gain appears in Long-Horizon Attribution, rising from 29.38\% to 36.08\% ($+$6.70).
For Qwen3, gains are largest in Long-Horizon Attribution, from 44.14\% to 56.38\% ($+$12.24), and similarly large in Spurious Co-occurrence, from 42.13\% to 52.34\% ($+$10.21).
These are the cases where fixed item-level confidence tends to blur who did what, when, and in which context.
Thus, MPRC is not merely correcting low-confidence items; it is strongest when real evidence must be assigned to the correct event, source, or distant context.

The improvement also scales with backbone quality.
Qwen3 starts from a stronger Raw~AV baseline than Qwen2.5, yet receives the larger MPRC gain.
This suggests that stronger backbones can expose cleaner confidence margins and more separable audio-counterfactual responses.
Rather than making calibration unnecessary, stronger perception supplies better reliability signals for stricter evidence-binding decisions.

\subsection{Ablation Study}

Table~\ref{tab:ablation} isolates components under a refined ablation protocol; the Full rows are diagnostic, not replacements for Table~\ref{tab:main_result}.
The w/o Aud. variant is a strong native-feature control: it uses the same reliability layer but removes audio-counterfactual branch features.
Adding those features improves SPA by 1.12 points for Qwen2.5 and 1.39 points for Qwen3, isolating the incremental value of binding-sensitive response shifts beyond native confidence calibration.
Score geometry, uncertainty, and compact selected branch counts support the design: MPRC benefits from structured probes without using all probes uniformly.

\vspace{0.25em}

\noindent
\begin{minipage}[t]{0.5\linewidth}
\vspace{0pt}
\centering
\tblfoot
\setlength{\tabcolsep}{2.7pt}
\renewcommand{\arraystretch}{1.25}
\begin{adjustbox}{width=\linewidth}
\begin{tabular}{@{}llrrr@{}}
\toprule
\rowcolor{tableHead}
\textbf{Model} & \textbf{Variant} & \textbf{All} & \textbf{Drop} & \textbf{K} \\
\midrule
\multirow{4}{*}{Qwen2.5}
& Full      & \textbf{36.22} & 0.00 & 2.60 \\
& w/o Geo.  & 34.67 & 1.55 & 1.40 \\
& w/o Unc.  & 35.05 & 1.17 & 2.80 \\
& w/o Aud.  & 35.10 & 1.12 & -- \\
\midrule
\multirow{4}{*}{Qwen3}
& Full      & \textbf{51.09} & 0.00 & 4.00 \\
& w/o Geo.  & 50.98 & 0.11 & 4.00 \\
& w/o Unc.  & 50.77 & 0.32 & 3.80 \\
& w/o Aud.  & 49.70 & 1.39 & -- \\
\bottomrule
\end{tabular}
\end{adjustbox}
\captionof{table}{Component ablation. Drop is relative to Full; Geo., Unc., and Aud. denote score geometry, uncertainty, and audio-counterfactual features.}
\label{tab:ablation}
\end{minipage}
\hfill
\begin{minipage}[t]{0.47\linewidth}
\vspace{0pt}
\centering
\includegraphics[width=0.96\linewidth]{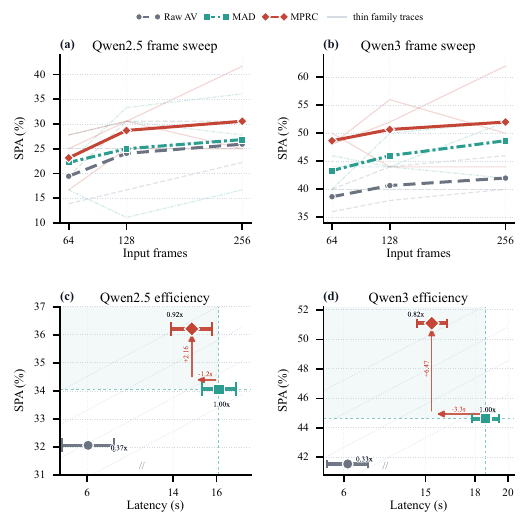}
\captionof{figure}{Frame-count and efficiency diagnostics. Panels compare Raw~AV, MAD, and \method under 64, 128, and 256 input frames and summarize latency versus SPA.}
\label{fig:frame_ablation}
\end{minipage}

\vspace{0.45em}

\subsection{Frame Count Analysis}

Figure~\ref{fig:frame_ablation} tests whether denser visual sampling or higher decoding cost can replace reliability calibration on per-model, subtype-balanced subsets.
Increasing frames improves Raw~AV, but \method remains higher at every density.
The efficiency view shows that selected-branch \method is faster than MAD while achieving higher SPA, because it evaluates only the training-selected probes.
Thus, broader visual coverage and heavier fixed decoding help, but neither resolves which recurring evidence should support the claim; evidence-binding reliability still requires calibration.

\vspace{0.30em}

\subsection{Cross-Benchmark Generalization}

\noindent
\begin{minipage}[t]{0.535\linewidth}
\vspace{0pt}
Table~\ref{tab:generalization} evaluates whether the reliability principle extends beyond paired Yes/No verification on OmniVideoBench and WorldSense.
These are not zero-shot transfer experiments.
For MCQ benchmarks, we do not verbalize each option as a binary claim; instead, \method calibrates choice-logit and audio-counterfactual option features under video-level target folds.
The Omni backbone remains frozen throughout.
This format shift tests whether audio-counterfactual reliability remains informative when the decision is expressed through competing answer options rather than hidden Yes/No pairs.
\end{minipage}
\hfill
\begin{minipage}[t]{0.43\linewidth}
\vspace{0pt}
\centering
\tblfoot
\setlength{\tabcolsep}{2.6pt}
\renewcommand{\arraystretch}{1.04}
\begin{adjustbox}{width=\linewidth}
\begin{tabular}{@{}lrrrr@{}}
\toprule
\rowcolor{tableHead}
\textbf{Model} & \textbf{Raw} & \textbf{\method} & \textbf{Cal.} & \textbf{Cal.+Opt.} \\
\midrule
\rowcolor{tableBand}\multicolumn{5}{@{}l}{\textit{OmniVideoBench}} \\
Qwen2.5-O-7B  & 34.30 & 35.50$_{+1.20}$ & 34.80 & 34.80$_{+0.50}$ \\
Qwen3-O-Inst. & 39.50 & 41.70$_{+2.20}$ & 40.80 & \textbf{43.10$_{+3.60}$} \\
\midrule
\rowcolor{tableBand}\multicolumn{5}{@{}l}{\textit{WorldSense}} \\
Qwen2.5-O-7B  & 43.00 & 45.24$_{+2.24}$ & 44.07 & \textbf{48.36$_{+5.36}$} \\
Qwen3-O-Inst. & 49.24 & 50.76$_{+1.51}$ & 50.32 & \textbf{54.16$_{+4.92}$} \\
\bottomrule
\end{tabular}
\end{adjustbox}
\captionof{table}{Cross-benchmark MCQ adaptation accuracy (\%). Subscripts show gains over Raw.}
\label{tab:generalization}
\end{minipage}

\vspace{0.25em}

The pure MPRC row improves all four target settings, with gains of $+$1.20 and $+$2.20 on OmniVideoBench and $+$2.24 and $+$1.51 on WorldSense.
Target calibration and option-position normalization further improve Qwen3 on OmniVideoBench ($+$3.60) and both models on WorldSense ($+$5.36 and $+$4.92).
Although these gains are smaller than in-domain OmniHalluc-L improvements, their consistent direction across benchmarks, backbones, and output formats suggests that MPRC captures a reusable reliability signal rather than a shortcut tied to OmniHalluc-L's pair format.
This is notable because the MCQ setting removes the hidden-pair structure entirely: the model must express reliability through competing option scores rather than a supported/counterfactual decision.
The positive direction therefore indicates that audio-counterfactual sensitivity is not merely a scoring artifact of OmniHalluc-L, but a portable signal for long-form Omni reliability.

\vspace{0.15em}

\section{Conclusion}
We presented OmniHalluc-L, a long-form Omni hallucination benchmark testing correct audio-visual evidence binding across speaker, moment, scene, and modality.
With 3{,}600 verified QA items from 1{,}800 pairs and 638 roughly 24-minute videos, OmniHalluc-L recasts hallucination evaluation as strict supported/counterfactual testing and reveals a reliability gap in open-weight Omni models.
We introduced MPRC, a frozen-backbone reliability layer calibrating AV confidence with audio-negative response shifts.
Gains on OmniHalluc-L and target-adapted MCQ benchmarks suggest a route to more reliable long-form Omni assistants.
We will release OmniHalluc-L and MPRC code.

\section*{Limitations}

\bench evaluates long-form evidence binding through single-claim verification.
This makes supported/counterfactual consistency directly measurable but does not cover multi-turn dialogue, open-ended generation, or interactive evidence seeking.
Extending strict binding evaluation to these settings remains future work.
\method is a post-hoc frozen-backbone calibration framework, most applicable to models exposing confidence signals or option scores; closed-source systems are therefore reported as Raw~AV references when these features are unavailable.

%% file: content/2_Appendix.tex
% If the main file already calls \appendix before inputting this file,
% do not call \appendix again.
\appendix

\section{Evaluation Protocol Details}
\label{sec:appendix_benchmark_details}
\label{sec:appendix_input_norm}
\label{sec:appendix_answer_normalization}

\paragraph{Input normalization.}
All benchmark statistics and model inputs are reported at the source-video level.
All evaluations use a fixed, query-agnostic preprocessing policy.
For models with explicit duration or context constraints, each source video is converted into the longest input representation accepted by the corresponding model interface.
For models without such constraints, we uniformly sample 256 frames from the source video.
This preprocessing is performed once per video and shared by all claims from that video.
It never uses evidence spans, claim-specific retrieval windows, pair identifiers, or paired counterfactual claims.

This protocol separates benchmark scale from model-interface constraints.
Dataset-level statistics describe the original long-form sources, while inference operates on normalized inputs determined only by the model interface.
The frame-count analysis in the main text is a controlled diagnostic over per-model, subtype-balanced subsets and should not be read as a replacement for the full-benchmark results.

\paragraph{Answer normalization.}
All benchmark items are evaluated as binary \textsc{Yes}/\textsc{No} claim verification.
For open-weight Omni models, we avoid free-form answer parsing whenever answer logits are available.
Under the same verification prompt, we read the logits of the canonical \textsc{Yes} and \textsc{No} alternatives and compute
\[
m_{\textsc{Yes}}=\ell_{\textsc{Yes}}-\ell_{\textsc{No}},
\qquad
p_{\textsc{Yes}}
=
\frac{\exp(\ell_{\textsc{Yes}})}
{\exp(\ell_{\textsc{Yes}})+\exp(\ell_{\textsc{No}})}
=
\sigma(m_{\textsc{Yes}}).
\]
Raw~AV uses this binary score for the native decision, while MAD and \method use the score, margin, and branch-induced changes as reliability features.

For systems that expose only generated text, we use the same \textsc{Yes}/\textsc{No} prompt and parse the response deterministically.
The parser normalizes case, whitespace, and punctuation, and accepts an answer only when the response contains an unambiguous \textsc{Yes} or \textsc{No} decision.
If a response is verbose, contains both labels, refuses to answer, or lacks a parseable binary decision, the same query is retried up to three times.
If all attempts remain unparseable, the item is marked incorrect.
No parser decision uses the paired counterfactual claim, evidence spans, or pair identity.
A hidden pair receives credit only when both claims produce parseable and correct binary decisions.

\paragraph{Closed-source references.}
Closed-source systems are reported only as Raw~AV references because token logits and confidence features are unavailable.
The public code release includes parser templates and configuration examples, while experiment-specific service identifiers and access-time metadata are kept in internal logs.

\paragraph{Implementation notes.}
All \textsc{Yes}/\textsc{No} verification runs use the same claim wrapper and score only the answer alternatives.
Model-specific input limits are handled by the fixed preprocessing rule above.
Efficiency diagnostics are reported within each backend rather than as cross-hardware speed comparisons.
Selected-$K$ \method timing sums selected branch calls and should be interpreted as an unfused implementation cost; it does not assume a shared or optimized multi-branch runner.

\section{Subtype Taxonomy}
\label{sec:appendix_subtype_taxonomy}

Table~\ref{tab:subtype_taxonomy} summarizes the relation-defined subtype taxonomy used in \bench.
The taxonomy is organized by the binding relation that must be preserved, not by surface video content.
This allows supported and counterfactual claims to share local evidence while differing only in how that evidence is connected.

\begin{table}[!tbp]
\centering
\tblsmall
\begin{adjustbox}{max width=\textwidth}
\begin{tabular}{@{}lll@{}}
\toprule
\rowcolor{tableHead}
\textbf{Family} & \textbf{Subtype} & \textbf{Binding relation tested} \\
\midrule
Cross-Stream Temporal
& TI: Temporal Inversion
& Event order across audio and visual streams \\
Cross-Stream Temporal
& BM: Boundary Misalignment
& Boundary alignment between speech, sound, and visual events \\
\midrule
Spurious Co-occurrence
& SM: Scene Misassociation
& False association between real evidence and the wrong scene \\
Spurious Co-occurrence
& SMB: Synchronous Misbinding
& Incorrect local binding among simultaneous cues \\
Spurious Co-occurrence
& CSS: Cross-Segment Splicing
& Spurious relation between evidence from different segments \\
\midrule
Long-Horizon Attribution
& CSSM: Speaker Misattribution
& Speaker attribution across distant moments \\
Long-Horizon Attribution
& CSIC: Identity Conflation
& Identity consistency across recurring appearances \\
Long-Horizon Attribution
& LRSC: Source Continuity
& Continuity of source, speaker, or event across the video \\
\bottomrule
\end{tabular}
\end{adjustbox}
\caption{Subtype taxonomy used in \bench. Subtypes are grouped by the evidence-binding relation they test.}
\label{tab:subtype_taxonomy}
\end{table}

\section{Verification Criteria}
\label{sec:appendix_verification}

The benchmark construction uses human verification to check whether each supported/counterfactual pair follows the intended strict-pair protocol.
A valid pair must satisfy three conditions:
the supported claim is entailed by the source evidence;
the counterfactual claim is plausible but false because of an incorrect binding relation;
and the two claims differ by the tested binding relation rather than by unrelated content, formatting artifacts, or answer-side cues.

Evidence intervals are retained only for construction and diagnostic use.
They are not provided to models during evaluation, calibration, or test-time inference.
The public repository does not include item-level annotations, evidence spans, source identifiers, labels, or pair membership.

\section{Sanity Checks}
\label{sec:appendix_textonly}

We run text-only sanity checks by removing video and audio evidence while keeping the same \textsc{Yes}/\textsc{No} verification format.
These probes are used only to diagnose language-prior shortcuts.
The primary Qwen Omni backbones perform far below the random-guessing strict-pair baseline under this setting, indicating that \bench is not solvable from claim text alone.
We omit the detailed text-only result table because it is diagnostic rather than part of the main comparison.

\section{Audio-Negative Probes}
\label{sec:appendix_audio_bank}
\label{sec:appendix_mad_scope}
\label{app:fixed_branch_diag}

\paragraph{Probe bank.}
\method uses audio perturbations as reliability probes rather than as replacement evidence.
Each branch keeps the visual stream fixed and perturbs only the audio stream, so response changes expose sensitivity to audio-visual binding rather than to visual perception.
The released code supports the compact branch bank in Table~\ref{tab:audio_probe_bank}.

\begin{table}[!tbp]
\centering
\tblsmall
\begin{adjustbox}{max width=0.88\textwidth}
\begin{tabular}{@{}lll@{}}
\toprule
\rowcolor{tableHead}
\textbf{Branch} & \textbf{Operation} & \textbf{Primary diagnostic role} \\
\midrule
\texttt{sh30}   & Global audio shift by 30s        & Light temporal misalignment \\
\texttt{sh60}   & Global audio shift by 60s        & Medium temporal misalignment \\
\texttt{seg60}  & Segment-level audio shift, 60s   & Local boundary disruption \\
\texttt{seg90}  & Segment-level audio shift, 90s   & Longer segment disruption \\
\texttt{swap60} & Adjacent audio segment swap, 60s & False local co-occurrence \\
\bottomrule
\end{tabular}
\end{adjustbox}
\caption{Audio-negative branch definitions for the released code configuration. Exact branch subsets are selected on training videos within each video-level fold.}
\label{tab:audio_probe_bank}
\end{table}

The branch bank is intentionally compact.
A broader perturbation pool is not automatically better, since destructive perturbations can remove evidence needed to judge the claim.
We therefore use structured audio-negative probes and select branch subsets on training videos within each fold.

\paragraph{MAD comparison scope.}
MAD is used as the closest frozen-decoding reference.
\method differs from MAD by fitting a lightweight reliability layer over native confidence and selected audio-negative response shifts, while keeping the Omni backbone frozen.
The comparison therefore asks whether evidence-binding reliability can be improved without updating the audio encoder, visual encoder, or language model.
For protocol consistency, MAD hyperparameters are selected only within training folds.

\paragraph{Fixed-branch diagnostics.}
We also check whether individual audio-negative branches expose useful reliability signal without fold-level selection.
These runs are diagnostic and are not reported as headline benchmark results.
The purpose is to verify that audio-negative views contain binding-sensitive information and to motivate learned branch selection instead of hard-coding one universal perturbation rule.

\section{External MCQ Adaptation}
\label{sec:appendix_external_mcq}

For external MCQ benchmarks, \method is adapted to option scoring rather than to binary supported/counterfactual pairs.
We do not rewrite each answer option into an OmniHalluc-L-style binary claim.
Instead, each option is represented by native audio-visual choice scores and by option-score shifts under audio-negative probes.
The feature map includes raw option scores, margins against competing options, and branch-induced score changes.
\method-Cal adds video-level calibration on the target benchmark, while Cal.+Opt. further includes option-position normalization.
This setup is a target-format adaptation rather than a zero-shot transfer claim: the Omni backbone remains frozen, and all target-benchmark adaptation is performed within video-level folds before evaluation on held-out videos.

\section{Prompt Templates}
\label{sec:appendix_prompts}

This section records only the task prompts needed to reproduce the public evaluation format.
Model-specific runner wrappers and system messages are kept in code/configuration files rather than repeated in the paper appendix.

\paragraph{\textsc{Yes}/\textsc{No} claim verification.}
\begin{quote}\small\ttfamily
Is the following statement about the video correct: \{query\} Answer only `Yes' or `No'. Do not include any explanation.
\end{quote}

\paragraph{MCQ direct decoding.}
\begin{quote}\small\ttfamily
Watch and listen to the video carefully. Answer the multiple-choice question using the video and audio evidence. Select exactly one option. Do not explain.\\[2pt]
Question: \{question\}\\
Options: (A) ... (B) ... (C) ... (D) ...\\
Final answer (A, B, C, D):
\end{quote}

\paragraph{MCQ option-score prompt.}
\begin{quote}\small\ttfamily
Watch and listen to the video carefully. Decide whether the candidate answer correctly answers the question using the video and audio evidence. Do not explain.\\[2pt]
Question: \{question\}\\
Candidate answer: \{option\_text\}\\
Final answer (Yes or No):
\end{quote}

\section{Public Code Release Scope}
\label{sec:appendix_release_scope}

The public repository contains the \method implementation, prompt templates, configuration examples, evaluation scripts, parser code, and documentation for running the protocol on user-provided data.
It does not include raw videos, source-video identifiers, item-level \bench annotations, evidence spans, labels, pair membership, model-output caches, or API-response logs.

This release scope supports method reproduction and evaluation-format reuse without distributing the benchmark annotation set.